\documentclass[lettersize,onecolumn,11pt]{IEEEtran}
\usepackage{amsmath,amssymb,amsfonts}
\usepackage{algorithmic}
\usepackage{algorithm}
\usepackage{array}
\usepackage[caption=false,font=normalsize,labelfont=sf,textfont=sf]{subfig}
\usepackage{textcomp}
\usepackage{stfloats}
\usepackage{url}
\usepackage{verbatim}
\usepackage{graphicx}
\usepackage{cite}

\usepackage{hyperref}
\hypersetup{hidelinks=true}

\newtheorem{theorem}{Theorem}

\newtheorem{lemma}{Lemma}
\newtheorem{corollary}{Corollary}
\newtheorem{definition}{Definition}
\newtheorem{remark}{Remark}
\newtheorem{assumption}{Assumption}

\begin{document}
	
	\title{Integrated Control and Communication \\in LQG Systems}
	
	\author{Sepehr Jahangiri, H. Ali Talebi,~\IEEEmembership{Senior Member,~IEEE}
		\thanks{S.~Jahangiri is with the Department of Electrical and Computer Engineering, University of Tehran, Tehran, Iran (e-mail: sjahangiri@ut.ac.ir). He was with the Department of Electrical Engineering, Amirkabir University of Technology, Tehran, Iran, when this work was carried out.}
		\thanks{H.~A.~Talebi is with the Department of Electrical Engineering, Amirkabir University of Technology, Tehran, Iran (e-mail: alit@aut.ac.ir).}}

	\maketitle
\begin{abstract}
	In this paper, we study the Integrated Communication and Control (ICAC) problem. Specifically, we investigate how messages can be transmitted from the controller/encoder to the observer/decoder through the control signal in Multiple-Input Multiple-Output (MIMO) vector-state Linear Quadratic Gaussian (LQG) systems under control constraints. We provide a computable capacity expression using semidefinite programming. We further show that it is possible to transmit data at a nonzero rate over an LQG system while maintaining the same optimal control cost as in the case where no information messages are transmitted. Finally, we discuss how this framework generalizes communication over MIMO Gaussian channels with feedback, both with and without InterSymbol Interference (ISI), so the expression also characterizes the capacity of these channels.
\end{abstract}

\begin{IEEEkeywords}
	Integrated communication and control, channel capacity, remote state estimation.
\end{IEEEkeywords}

\section{Introduction} \label{sec:0}

Control and communication are two closely related fields, and their interaction appears in many problems. A common setting is controlling a dynamic system through communication links under communication constraints \cite{tati04,tati04stoch,khina19,sahai1,zaidi,kostina}. Another line of work focuses on designing communication schemes that work well with control tasks. There is also a long history of studying the connection between controlling a dynamic system through a channel and sending messages over a channel with feedback \cite{sahai1,bode}. The well-known Schalkwijk–Kailath (SK) feedback coding scheme \cite{SK} is a classic example. Many works have extended or analyzed this scheme under different assumptions \cite{SK11,SK12,SK13,SK2}.

\begin{figure}[t!]
	\centering
	\includegraphics[width=6.5in]{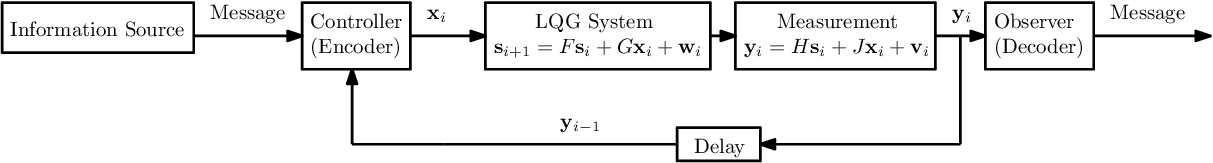}
	\caption{Schematic of Integrated Control and Communication (ICAC) in LQG Systems.}
	\label{fig:1}
\end{figure}

There is a different type of interaction between these fields, which we refer to as Integrated Control and Communication (ICAC). Here, the controller of a dynamic system sends information to an observer, but there is no dedicated communication link. Instead, the controller embeds messages into the control inputs, while still controlling the system. The observer only sees the system outputs and must recover the messages from them. A diagram of this setup appears in Fig.~\ref{fig:1}. We also know that communication over a MIMO Gaussian channel with memory, with or without Inter Symbol Interference (ISI), can be seen as communication over a dynamic system \cite{sabagMemory}, so ICAC is a general case of feedback communication over MIMO Gaussian channels with memory.

ICAC has a wide range of potential applications. For example, in multi-agent systems, a leader may need to transmit information to other agents, while in autonomous driving, vehicles may need to communicate their intentions, such as lane changes, to nearby cars \cite{ChenICAC1,ChenICAC2}. These interactions can be achieved without an explicit communication link, solely using the dynamical system as an implicit communication channel. Even if an explicit communication link is available, this framework can still be utilized. Another potential application arises in V2X communications: when a base transceiver station (BTS) employs ISAC signaling for downlink transmission, ICAC can enable V2X communication without requiring an explicit uplink communication link \cite{ISAC}.

In this work, we study ICAC over linear discrete time systems with an LQR cost. We consider the LQG case where neither the controller nor the observer has access to the actual state; both only observe noisy outputs. Our goal is to compute the communication capacity of this setup.
\subsection{Prior Works}

There is already a large body of work related to this topic. The feedback capacity of communication channels and its connection to directed information have been studied in \cite{dirC1,kramer1998directed}. The feedback capacity of MIMO Gaussian channels, which can be seen as a simple form of ICAC with an input cost, was first given in \cite{cover}, but the expression was not computable. A computable formula was later provided in \cite{sabagfeedback}. The same authors studied MIMO Gaussian channels with ISI in \cite{sabagMemory}, where they derived an upper bound for capacity and conjectured that this bound is exact, proving it for the scalar case. As a result of our analysis of ICAC, we also obtain the feedback capacity of MIMO Gaussian channels with ISI and show that their conjecture is correct. Other channel models with known feedback capacity include \cite{FB1,FB2}. For Markov channels with state feedback, computable capacity expressions exist \cite{mark1,mark2,mark3}, but extending these ideas to measurement feedback is more difficult.

ICAC scenarios when the state, input, and output take values in discrete alphabets has been studied in \cite{ChenICAC1,discrete1,discrete2}. In Gaussian dynamic systems where both the controller and observer have full access to the state, ICAC capacity results were also obtained in \cite{discrete1,discrete2}. For the measurement-feedback case, where both sides only observe the system outputs, $n$-letter characterizations were proposed in \cite{chara1,chara2}, but these expressions are not computable since their proposed coding policy is time varying.

The authors of \cite{ChenICAC2} analyzed two special cases in which either the controller or the observer sees the full system state while the other only sees partial measurements. They derived the ICAC capacity for these cases and gave a lower bound for the case where both rely on output measurements.

More recently, the authors in \cite{sabagICAC} studied the ICAC capacity of LQG systems with measurement feedback. They derived a computable upper bound and gave conditions under which the bound is tight and achievable using a time-invariant coding scheme. They also proposed a conjecture, similar to the one in \cite{sabagMemory}, that the capacity always equals to this bound, and also provided numerical evidence. With additional assumptions, they also computed the capacity for single-state SISO LQG systems.

\subsection{Contribution}
In this paper, we provide a computable convex optimization method, based on semidefinite programming, for computing the ICAC capacity of general multi-state MIMO LQG systems under LQR constraints. To this end, we show that the conjecture in \cite{sabagICAC} holds, meaning that the capacity equals the proposed upper bound.

Since ICAC over LQG systems with LQR constraints generalizes feedback communication over MIMO Gaussian channels, our capacity formula also applies to those channels; thus, the main theorem also resolves the conjecture in \cite{sabagMemory} for the capacity of MIMO Gaussian channels with memory.

We also show that a nonzero rate can be achieved with the same LQR cost (asymptotically) as optimal control without communication.

Sections~\ref{sec:2} and~\ref{sec:3} contain the main contributions of the paper.

\subsection{Paper Structure}
The rest of the paper is organized as follows. Section~\ref{sec:1} presents the problem setup, preliminaries, and results from prior works. Section~\ref{sec:2} presents our main results and contributions, and Section~\ref{sec:3} provides the proof of the main result. Proofs and discussions of supporting lemmas appear in the appendices.

\section{Setting and Preliminaries}\label{sec:1}
In this section, we first introduce the notations and definitions used throughout the paper. We then describe the problem setup, discuss some basic reductions for LQG systems, and present the ICAC formulation.
\subsection{Notations and Definitions}\label{sec:1-1}
Random vectors are denoted by bold lowercase letters, such as $\mathbf{x}$, and deterministic matrices are denoted by uppercase letters, such as $F$. Superscript indices are used as $\mathbf{x}^{i} = [\mathbf{x}_1, \ldots, \mathbf{x}_i]$. The symbol $\lambda_i(A)$ denotes the $i$-th largest eigenvalue of the matrix $A$. The matrix $A^\dagger$ represents the Moore–Penrose pseudoinverse of $A$. We use $\bar{\mathbf{x}}$ to denote the complex conjugate of $\mathbf{x}$. A matrix $A$ is called stable if all its eigenvalues lie strictly inside the unit circle. Finally $\mathcal{S}(A)$ is defined as the Schur complement of a matrix $A$.

Below are several definitions that we use in the paper.
\begin{definition}
	\textit{The pair $(F, H)$ is detectable, if for any $x$ and $\lambda$ such that $Fx=\lambda x$ if $|\lambda|\ge 1$, then $Hx\neq 0$ .}
\end{definition}
\begin{definition}
	\textit{ The pair $(F, W)$ is stabilizable, if for any $x$ and $\lambda$ such that $x^TF=x^T\lambda$ if $|\lambda|\ge 1$, then $x^TW\neq 0$ .}
\end{definition}
\begin{definition}
	\textit{The pair $(F, W)$ is controllable on the unit circle, if for any $x$ and $\lambda$ such that $x^TF=x^T\lambda$ if $|\lambda|=1$ , then $x^TW\neq 0$ .}
\end{definition}

To help the reader keep track of the many symbols used in this paper, Table~\ref{tab:notation} summarizes the main quantities. For each symbol, the table gives its meaning and the equation or section where it is first defined. Since the table refers to equations introduced later in Sections~\ref{sec:1-2} to~\ref{sec:3}, it is intended as a reference to be consulted while reading, rather than to be read in full at this point.

\begin{table}[t]
	\caption{Summary of main symbols}\label{tab:notation}
	\centering
	\renewcommand{\arraystretch}{1.15}
	\begin{tabular}{@{}l p{0.62\linewidth} l@{}}
		\hline
		\textbf{Symbol} & \textbf{Meaning} & \textbf{Defined}\\
		\hline
		\multicolumn{3}{@{}l}{\emph{System}}\\
		$\mathbf{s}_i,\mathbf{x}_i,\mathbf{y}_i$ & state, input, output ($\in\mathbb{R}^r,\mathbb{R}^p,\mathbb{R}^l$) & \eqref{eq:1}\\
		$F,G,H,J$ & state, input, output, feedthrough matrices & \eqref{eq:1}\\
		\multicolumn{3}{@{}l}{\emph{Estimation}}\\
		$\hat{\mathbf{s}}_i$ & Kalman estimate of $\mathbf{s}_i$ at the encoder & \eqref{eq:2}\\
		$\mathbf{e}_i,\Psi$ & innovation process and its covariance & \eqref{eq:6},\eqref{eq:8}\\
		$\Sigma,K_p$ & steady-state Kalman error covariance and gain & \eqref{eq:7},\eqref{eq:8}\\
		$\hat{\hat{\mathbf{s}}}_i$ & decoder's MMSE estimate of $\hat{\mathbf{s}}_i$ given $\mathbf{y}^{i-1}$ & \eqref{eq:17.5}\\
		\multicolumn{3}{@{}l}{\emph{LQR}}\\
		$Q,R$ & state and input cost weights & \eqref{eq:9}\\
		$E,K_{\mathtt{LQR}},\Psi_{\mathtt{LQR}}$ & solution, gain, and input-weight term of the control Riccati equation & \eqref{eq:12}\\
		$\mathcal{J}^*$ & optimal cost without communication & \eqref{eq:13}\\
		$p$ & LQR cost budget & Sec.~\ref{sec:1-3}\\
		\multicolumn{3}{@{}l}{\emph{Coding policy and optimization}}\\
		$\hat\Sigma$ & steady-state covariance of $\hat{\mathbf{s}}_i-\hat{\hat{\mathbf{s}}}_i$ (design variable) & \eqref{eq:14}\\
		$\Pi$ & covariance of the input component & \eqref{eq:14}\\
		$\Gamma$ & cross-covariance between input and $\hat{\mathbf{s}}_i-\hat{\hat{\mathbf{s}}}_i$ & \eqref{eq:14}\\
		$M=\Pi-\Gamma\hat\Sigma^\dagger\Gamma^T$ & covariance of the i.i.d.\ part $\mathbf{m}_i$ & \eqref{eq:17}\\
		$\Psi_y,K_y$ & covariance and gain of the decoder innovation & \eqref{eq:14c4},\eqref{eq:14c5}\\
		$\Upsilon$ & stacked variable $\begin{bmatrix}\hat\Sigma&\Gamma^T\\\Gamma&\Pi\end{bmatrix}$ & \eqref{eq:20}\\
		$\varepsilon$ & margin enforcing $M\succeq\varepsilon I$ & \eqref{eq:ach-a}\\
		$C_{\mathtt{UB}},C_{\mathtt{LB},\varepsilon}$ & upper bound \eqref{eq:14} and $\varepsilon$-lower bound \eqref{eq:ach} & Sec.~\ref{sec:3}\\
		\multicolumn{3}{@{}l}{\emph{Proof objects}}\\
		$\tilde\Sigma_i,\tilde\Psi_{y,i},\tilde K_{y,i}$ & time-varying Riccati recursion of the scheme & \eqref{eq:ll1}--\eqref{eq:ll3}\\
		$\mathcal{O}_\varepsilon$ & optimizer set of \eqref{eq:ach} & Sec.~\ref{sec:3}\\
		\hline
	\end{tabular}
\end{table}
\subsection{LQG Systems} \label{sec:1-2}

The linear dynamic system is described by the state–space model
\begin{align} 
	\label{eq:1}
	\mathbf{s}_{i+1}&=F\mathbf{s}_i+G\mathbf{x}_i+\mathbf{w}_i \nonumber \\
	\mathbf{y}_i&=H\mathbf{s}_i+J\mathbf{x}_i+\mathbf{v}_i
\end{align}
where $\mathbf{x}_i \in \mathbb{R}^p$ is the control input, $\mathbf{y}_i \in \mathbb{R}^l$ is the system output (or measurement), and $\mathbf{s}_i \in \mathbb{R}^r$ is the system state. The disturbances are given by $\mathbf{w}_i \sim \mathcal{N}(0, W)$ and the measurement noises by $\mathbf{v}_i \sim \mathcal{N}(0, V)$, where both are i.i.d. sequences with cross-covariance $\mathbb{E}[\mathbf{w}_i \mathbf{v}_i^T] = L$. They are also independent of the initial state $\mathbf{s}_1 \sim \mathcal{N}(0, \Sigma_1)$. The matrices $F$, $G$, $H$, and $J$ are known to both the encoder/controller and the decoder/observer.

The encoder/controller receives causal feedback of the system outputs, so each input $\mathbf{x}_i$ is a function of the message and the past outputs $\mathbf{y}^{i-1}$. It is also worth noting that even if the encoder only has delayed feedback, this delay can be removed through state augmentation. By adding extra state variables to capture the delayed outputs or inputs, the system can be converted into an equivalent delay–free model, following a similar approach to \cite{sabagMemory}.

Since this is a measurement-feedback system, neither the encoder nor the decoder has direct access to the true system states. To handle this, we convert the model into an equivalent state-space form where the encoder does have access to a state estimate, obtained through a Kalman filter. We define the estimated state as

\begin{equation}
	\label{eq:2}
	\hat{\mathbf{s}}_i=\mathbb{E}[\mathbf{s}_i|\mathbf{x}^{i-1},\mathbf{y}^{i-1}] 
\end{equation}
which can be computed recursively using
\begin{equation}
	\label{eq:3}
	\hat{\mathbf{s}}_{i+1}=F\hat{\mathbf{s}}_i+G\mathbf{x}_i+K_{p,i}(\mathbf{y}_i-J\mathbf{x}_i-H	\hat{\mathbf{s}}_i)
\end{equation}
with the initial condition $\hat{\mathbf{s}}_1 = 0$. The estimation error covariance, $\Sigma_i=\operatorname{Cov}(\hat{\mathbf{s}}_i-{\mathbf{s}}_i)$ is updated using

\begin{equation}
	\label{eq:4}
	\Sigma_{i+1}=F\Sigma_i F^T+W-K_{p,i}\Psi_i K_{p,i}^T
\end{equation}
where $\Sigma_1$ is the initial covariance and the constants are $K_{p,i}=(F\Sigma_i H^T+L)\Psi_i^{-1}$, and $\Psi_i=H\Sigma_i H^T+V$.

The new state–space model, in which the encoder has access to the estimated states, is given by
\begin{align}
	\label{eq:5}
	\hat{\mathbf{s}}_{i+1}&=F\hat{\mathbf{s}}_i+G\mathbf{x}_i+K_p\mathbf{e}_i \nonumber \\
	\mathbf{y}_i&=H\hat{\mathbf{s}}_i+J\mathbf{x}_i+\mathbf{e}_i
\end{align}
where the measurement noise term in this state-space is
\begin{equation}
	\label{eq:6}
	\mathbf{e}_i=\mathbf{y}_i-J\mathbf{x}_i-H	\hat{\mathbf{s}}_i.
\end{equation}
This process $\mathbf{e}_i$ is called the innovation process. It is white, Gaussian with distribution $\mathbf{e}_i \sim \mathcal{N}(0, \Psi_i)$, and is independent of $(\mathbf{x}^i, \mathbf{y}^{i-1})$ \cite{linearestimation}.

Let $\Sigma$ denote the maximal solution of the Riccati equation
\begin{equation}
	\label{eq:7}
	\Sigma=F\Sigma F^T+W-K_p\Psi K_p^T
\end{equation}
where
\begin{align}
	\label{eq:8}
	K_p&=(F\Sigma H^T+L)\Psi^{-1} \nonumber \\
	\Psi&=H\Sigma H^T+V.
\end{align}
Below, we provide a sufficient condition under which the sequence $\Sigma_i$ converges to the steady-state solution $\Sigma$.

\begin{assumption} \label{asmp:det}
	\textit{The pair $(F,H)$ is detectable.}
\end{assumption}

\begin{assumption} \label{asmp:cont}
	\textit{The pair $(F_s,W_s^{\frac{1}{2}})$ is controllable on the unit circle. Where $F_s=F-LV^{-1}H$ and $W_s=W-LV^{-1}L^T$.}
\end{assumption}

Assumptions~\ref{asmp:det} and~\ref{asmp:cont} are common in LQG scenarios to ensure that the Riccati equation for Kalman filtering has Stabilizing solution \cite{linearestimation}. We further assume that the covariance matrix converges to the stabilizing solution. There are several settings that can result in this convergence, here we consider Assumption 3. However, there are other settings as well, for example, a main result of \cite{dare}, which is also used later in the proof of our main results.

\begin{assumption} \label{asmp:stabKF}
	\textit{The pair $(F_s,W_s^{\frac{1}{2}})$ is stabilizable.}
\end{assumption}

We know that with these assumption $\Sigma$ converge to the unique stabilizing solution and for this solution we know that $F-K_pH$ is stable.

For any control policy, the finite-horizon LQR cost is
\begin{equation}
	\label{eq:9}
	\mathcal{J}_n=\frac{1}{n}[\mathbb{E}[\hat{\mathbf{s}}_{n+1}^TQ\hat{\mathbf{s}}_{n+1} ]+\sum_{i=1}^{n}(\mathbb{E}[\hat{\mathbf{s}}_{i}^TQ\hat{\mathbf{s}}_{i}]+\mathbb{E}[\mathbf{x}_i^TR\mathbf{x}_i])]+\frac{n+1}{n}\operatorname{Tr}(\Sigma Q).
\end{equation}
The infinite-horizon LQR cost is defined as
\begin{equation}
	\label{eq:10}
	\mathcal{J}=\lim\limits_{n\to\infty}\mathcal{J}_n.
\end{equation}
The optimal control policy that minimizes this cost is
\begin{equation}
	\label{eq:11}
	\mathbf{x}_i=-K_\mathtt{LQR}\hat{\mathbf{s}}_i
\end{equation}
where $E$, $K_{\mathtt{LQR}}$, and $\Psi_{\mathtt{LQR}}$ are the stabilizing solution of the Riccati equation
\begin{align}\label{eq:12}
	E&=F^T E F+Q-K_\mathtt{LQR}^T\Psi_\mathtt{LQR}K_\mathtt{LQR} \nonumber \\
	K_\mathtt{LQR}&=\Psi_\mathtt{LQR}^{-1}G^TEF \nonumber \\
	\Psi_\mathtt{LQR}&=R+G^TEG.
\end{align}
The optimal infinite-horizon LQR cost is then
\begin{equation}\label{eq:13}
	\mathcal{J}^*=\operatorname{Tr}(K_p\Psi K_p^T E)+\operatorname{Tr}(\Sigma Q).
\end{equation}

In the next part, we provide sufficient conditions for the convergence of the Riccati recursion to the unique stabilizing solution of \eqref{eq:12}.

\begin{assumption} \label{asmp:stab}
	\textit{The pair $(F,G)$ is stabilizable, $Q\succeq0$ and $R\succ 0$.}
\end{assumption}

Assumption~\ref{asmp:stab} ensures that the Riccati recursion converges to a steady-state matrix $E$, which determines the optimal feedback gain $K_{\mathtt{LQR}}$ for the infinite-horizon problem, however, it does not guarantee that it converges to a stabilizing solution  \cite{linearestimation}. It is worth noting that if $Q = 0$, meaning there is no penalty on the system state, controllability is not required, and the steady-state values become $E = 0$, $K_{\mathtt{LQR}} = 0$, and $\Psi_{\mathtt{LQR}} = R$.

\begin{assumption}\label{asmp:nonzero} \textit{
		The transfer function from the input to the output,
		\begin{equation}
			\tilde T(z) = J + H(zI-F)^{-1}G,
		\end{equation}
		is not identically zero as a rational matrix function, i.e.\ $\tilde T\not\equiv 0$.}
\end{assumption}

Assumption~\ref{asmp:nonzero} is a nondegeneracy condition that
ensures the input can influence the output at all. If $\tilde T\equiv 0$, the output
$\mathbf{y}_i$ is statistically independent of the input sequence, so no message
embedded in $\mathbf{x}^n$ can be recovered by the observer. In that case
$I(\mathbf{x}^n\to\mathbf{y}^n)=0$ and $C_{\mathtt{LQG}}(p)=0$ for every $p$, and the
problem is trivial. Assumption~\ref{asmp:nonzero} therefore only excludes this degenerate
case, and it is satisfied by essentially every system of interest.

\begin{assumption} \label{asmp:LQR}
	\textit{The pair $(F - G K_{\mathtt{LQR}}, H - JK_{\mathtt{LQR}})$ is detectable.}
\end{assumption}

Assumption~\ref{asmp:LQR} ensures that the feedback preserves the detectability property. This condition depends on $Q$ and $R$. For example, in the case of $Q = 0$, where there is no penalty on the system state (e.g., in feedback communication over MIMO Gaussian channels with memory), we have $K_{\mathtt{LQR}} = 0$, and this assumption holds. Alternatively, in the case where $(F - G K_{\mathtt{LQR}})$ is stable and $K_{\mathtt{LQR}}$ is the stabilizing solution to the Riccati equations, this assumption also holds.

\subsection{ICAC in LQG Systems} \label{sec:1-3}
Integrated Control and Communication (ICAC) refers to the problem of transmitting a message to an observer through the control input, while ensuring a given LQR cost. In other words, we aim to control the system and send information simultaneously.

For a fixed blocklength $n$ and a rate $\mathcal{R}$, the encoder (controller) sets the system input as a function of the message $\mathcal{M}\sim U([1:2^{n\mathcal{R}}])$ and previous system outputs, and the decoder (observer) has access to system measurements and estimates the message as $\hat{\mathcal{M}}$. We say that a rate $\mathcal{R}$ and LQR cost $p$ are achievable if there exists a coding policy such that the probability of error $\mathbb{P}(\mathcal{M}\neq \hat{\mathcal{M}})$ vanishes as $n$ grows. The ICAC capacity, given an LQR constraint $p$, is defined as the supremum over all rates $\mathcal{R}$ such that $(\mathcal{R}, p)$ is achievable.

The ICAC capacity is an extension of the feedback capacity of MIMO Gaussian channels with memory, which can be expressed using the following $n$-letter optimization:
\begin{equation}
	C_{\mathtt{LQG}}(p)=\lim\limits_{n\to\infty} \frac{1}{n} \max_{P(\mathbf{x}^n||\mathbf{y}^n),\mathcal{J}_n\le p} I(\mathbf{x}^n\to\mathbf{y}^n),
\end{equation}
where $I(\mathbf{x}^n\to\mathbf{y}^n)$ is the directed information, and $P(\mathbf{x}^n||\mathbf{y}^n)$ is the causal conditioning \cite{dirC1, kramer1998directed, dirC2}.

The authors in \cite{sabagICAC} derived an upper bound on the rate at which messages can be reliably transmitted over the system \eqref{eq:1}, subject to the constraint that the LQG cost does not exceed a given value $p$. They showed that this upper bound can be expressed as a convex optimization problem:
\begin{equation} \label{eq:14}
	C_{\mathtt{LQG}}(p)\le \max_{\Gamma,\Pi,\hat{\Sigma}}\frac{1}{2}\log(\frac{\det(\Psi_y)}{\det(\Psi)})
\end{equation}
\begin{subequations}
	\begin{equation}
		\mathrm{s.t.}\;\operatorname{Tr}(\hat{\Sigma}K_\mathtt{LQR}^T\Psi_\mathtt{LQR}K_\mathtt{LQR})+\operatorname{Tr}(\Pi\Psi_\mathtt{LQR})+2\operatorname{Tr}(\Gamma K_\mathtt{LQR}^T\Psi_\mathtt{LQR})+\mathcal{J}^*\le p, \label{eq:14c1}
	\end{equation}
	\begin{equation}
		\begin{bmatrix}
			\hat{\Sigma} & \Gamma^T \\ 
			\Gamma & \Pi
		\end{bmatrix}\succeq 0,\label{eq:14c2}
	\end{equation}
	\begin{equation}
		\begin{bmatrix}
			F\hat{\Sigma}F^T+G\Pi G^T+G\Gamma F^T+F \Gamma^T G^T+K_p\Psi K_p^T-\hat{\Sigma} & K_y\Psi_y \\ 
			\Psi_y K_y^T & \Psi_y
		\end{bmatrix}\succeq 0, \label{eq:14c3} 
	\end{equation}
	\begin{equation}
		\Psi_y=H\hat{\Sigma}H^T+J\Pi J^T+H\Gamma^T J^T+J \Gamma H^T+\Psi, \label{eq:14c4} 
	\end{equation}
	\begin{equation}
		K_y=(F\hat{\Sigma}H^T+F\Gamma^T J^T+G\Gamma H^T+G\Pi J^T+K_p \Psi)\Psi_y^{-1} \label{eq:14c5}
	\end{equation}
\end{subequations}
where $(K_p,\Psi,\Sigma)$ are the Kalman filter constants from \eqref{eq:7} and \eqref{eq:8}, and $(K_\mathtt{LQR},\Psi_\mathtt{LQR},E)$ are the LQR constants from \eqref{eq:12}. 

A MIMO Gaussian channel with memory and ISI under an average power constraint is a special case of ICAC, where in the LQR cost we have $Q=0$ and $R=I$ \cite{sabagMemory}. In \cite{sabagMemory}, it was shown that the feedback capacity of MIMO Gaussian channels with memory and ISI is upper bounded by the convex optimization problem 
\begin{equation}
	C_{\mathtt{LQG}}(p)\le \max_{\substack{\Gamma,\Pi,\hat{\Sigma}\\\eqref{eq:14c2},\eqref{eq:14c3},\eqref{eq:14c4},\eqref{eq:14c5}}}\frac{1}{2}\log(\frac{\det(\Psi_y)}{\det(\Psi)}) \label{eq:19}
\end{equation}
\begin{equation}
	\mathrm{s.t.}\;\operatorname{Tr}(\Pi)\le p \label{eq:19c}
\end{equation}

The authors of \cite{sabagMemory}, which considers MIMO Gaussian channels with memory and ISI, and \cite{sabagICAC}, which addresses the general ICAC setting, conjecture that  upper bound \eqref{eq:14} is achievable under the following time-invariant policy:
\begin{equation}\label{eq:17}
	\mathbf{x}_i=-K_{\mathtt{LQR}}\hat{\hat{\mathbf{s}}}_i+\Gamma\hat{\Sigma}^{\dagger}(\hat{\mathbf{s}}_i-\hat{\hat{\mathbf{s}}}_i)+\mathbf{m}_i.
\end{equation}

In this policy, the additive term $\mathbf{m}_i \sim \mathcal{N}(0,\Pi-\Gamma\hat{\Sigma}^\dagger \Gamma^T)$ is independent of $(\mathbf{m}^{i-1},\mathbf{x}^{i-1},\mathbf{y}^{i-1})$ for $(\Gamma,\Pi,\hat{\Sigma})$ being the optimal solution of \eqref{eq:14} that meets the sufficient conditions. The term $\hat{\hat{\mathbf{s}}}_i$ is the Minimum Mean Square Error (MMSE) estimate of $\hat{\mathbf{s}}_i$ based on the past outputs $\mathbf{y}^{i-1}$, computed at the decoder using Kalman filtering.
\begin{equation}
	\label{eq:17.5}
	\hat{\hat{\mathbf{s}}}_{i+1}=(F-GK_{\mathtt{LQR}})\hat{\hat{\mathbf{s}}}_{i}+K_{y,i}(\mathbf{y}_i-(H-JK_{\mathtt{LQR}})\hat{\hat{\mathbf{s}}}_{i}).
\end{equation}

Also note that, based on the constraints in \eqref{eq:14} and by using the Schur complement, we can confirm that $\Pi-\Gamma\hat{\Sigma}^\dagger \Gamma^T$, which is the covariance of the i.i.d. part, is positive semidefinite. 

The intuition behind this conjecture, which states that the proposed coding scheme can achieve the upper bound in \eqref{eq:14}, comes from the results of \cite{sabagfeedback} on the capacity of MIMO Gaussian channels with feedback and without ISI. The authors in \cite{sabagMemory} and \cite{sabagICAC} have also proved the conjecture for some special cases (e.g., scalar channels with a single state in \cite{sabagICAC}) and provided some intuition for why it may hold in the general case. However, to the best of our knowledge, this conjecture has not been proved in the general case, and no general result on the capacity of these channels is available in the literature.

In Section~\ref{sec:2}, we derive a computable expression for the ICAC capacity by proving this conjecture.

\section{Main Result} \label{sec:2}
In this section, we compute the capacity of any MIMO LQG system under an LQR cost constraint. We show that the capacity of the LQG system is equal to the upper bound in \eqref{eq:14}. In Section \ref{sec:2-2}, we also provide an example to show how these properties behave in a dynamic system. One interesting result is that we can achieve the same LQR cost as the optimal control policy while still being able to send information at a non-zero rate.

\subsection{Capacity of ICAC in LQG Systems}\label{sec:2-1}
Here we present a convex optimization problem that computes the capacity of LQG systems under LQR cost constraints. This capacity represents the highest amount of information that can be sent through a dynamic system while it is being controlled at the same time. We also recall that dynamic systems with an LQR cost are a general form of MIMO Gaussian channels with power constraints.

\begin{theorem}\label{thm:1}
	\textit{The capacity of an LQG system with LQR cost constraint $p>\mathcal{J}^*$ is given by the convex optimization
		\begin{equation}
			C_{\mathtt{LQG}}(p)= \max_{\substack{\Gamma,\Pi,\hat{\Sigma}\\\eqref{eq:14c1},\eqref{eq:14c2},\eqref{eq:14c3},\eqref{eq:14c4},\eqref{eq:14c5}}}\frac{1}{2}\log(\frac{\det(\Psi_y)}{\det(\Psi)})  \label{eq:29}
	\end{equation}} 
\end{theorem}

It is also worth mentioning again that this optimization is convex because the objective function is concave in $\Psi_{y}$, and $\Psi_{y}$ is a linear function of $(\Gamma,\Pi,\hat{\Sigma})$ with Linear Matrix Inequalities (LMIs). The capacities studied in \cite{sabagICAC,sabagfeedback,sabagMemory} are all special cases of Theorem~\ref{thm:1}.

Using Theorem~\ref{thm:1}, we can compute the feedback capacity of MIMO Gaussian channels with colored noise, with or without ISI, and more generally any LQG system with an LQR cost constraint. To compute the feedback capacity of MIMO Gaussian channels with memory and ISI under a power constraint, we simply set $Q=0$ and $R=I$. The following is the corollary of Theorem~\ref{thm:1} for the capacity of MIMO Gaussian channels with memory and ISI.

\begin{corollary}
	\textit{The feedback capacity of MIMO Gaussian channels with colored noise and ISI under a power constraint $p$ is given by the convex optimization
		\begin{equation}
			C_{\mathtt{LQG}}(p)= \max_{\substack{\Gamma,\Pi,\hat{\Sigma}\\\eqref{eq:14c2},\eqref{eq:14c3},\eqref{eq:14c4},\eqref{eq:14c5},\eqref{eq:19c}}}\frac{1}{2}\log(\frac{\det(\Psi_y)}{\det(\Psi)}) 
		\end{equation} } 
\end{corollary}

\begin{remark}
	\textit{It follows directly from \eqref{eq:14c1} that the capacity is zero for $p<\mathcal{J}^*$, since $\mathcal{J}^*$ is the minimum LQR cost achievable even when no information is transmitted. Thus, transmitting information cannot result in a lower cost. Theorem~\ref{thm:1} is therefore valid only for $p>\mathcal{J}^*$, while the case $p=\mathcal{J}^*$ is not studied in this paper. Nevertheless, throughout the paper, we say that a transmission rate $\mathcal{R}$ is achievable while maintaining the optimal LQR cost if	$\mathcal{R}<\lim_{p\to(\mathcal{J}^*)^+}C_{\mathtt{LQG}}(p),$ which means that the LQR cost can be made arbitrarily close to its optimal value.}
\end{remark}

\begin{figure}[b!]
	\centering
	\includegraphics[width=4.5in]{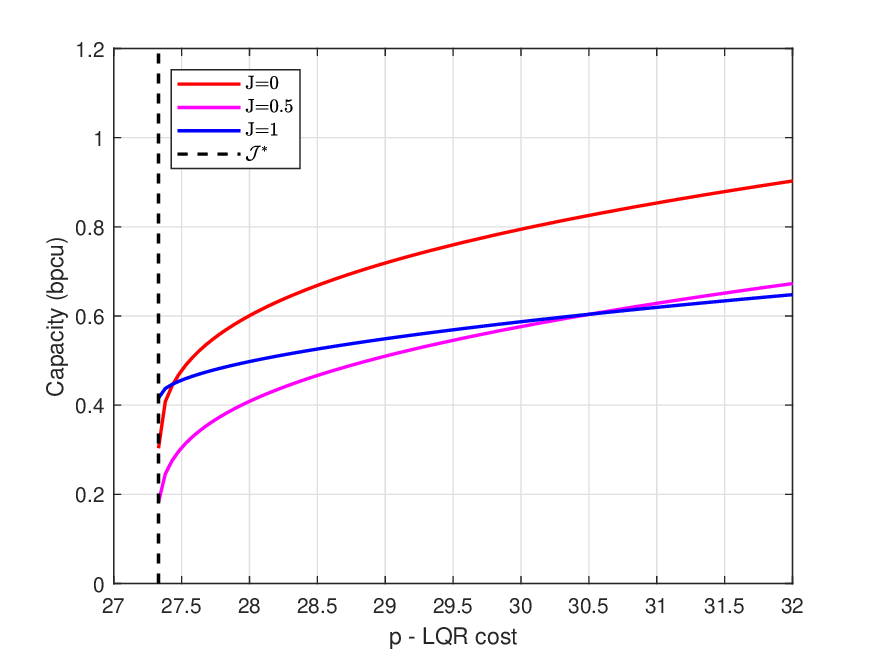}
	\caption{ICAC capacity of LQG systems \eqref{eq:30} for 
		\( J \in \{0, 0.5, 1\} \). The dashed line illustrates the minimum achievable LQR cost.}
	\label{fig:2}
\end{figure}
\begin{figure}[b!]
	\centering
	\subfloat[]{\includegraphics[width=2.5in]{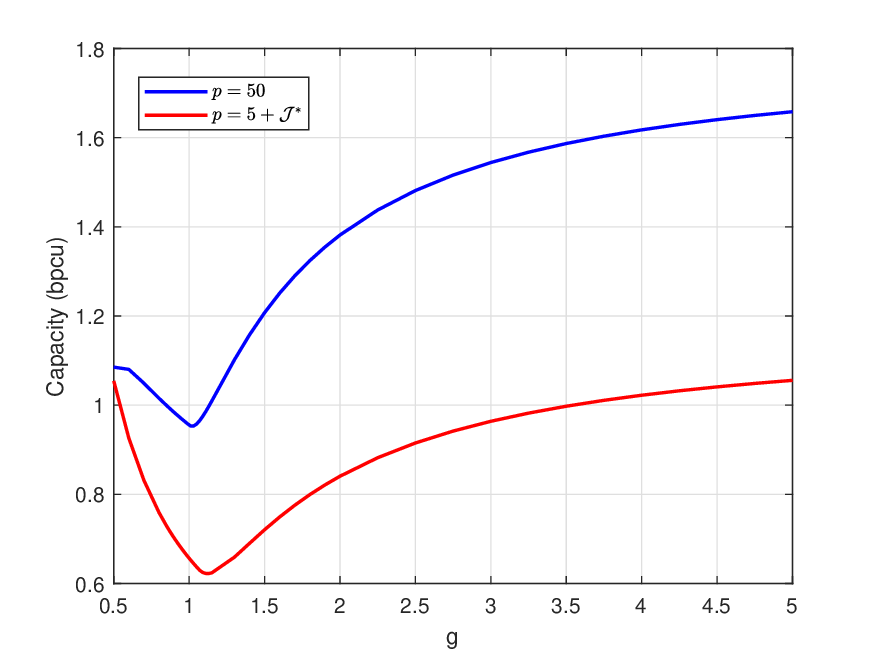}%
		\label{fig:3a}}
	\hfil
	\subfloat[]{\includegraphics[width=2.5in]{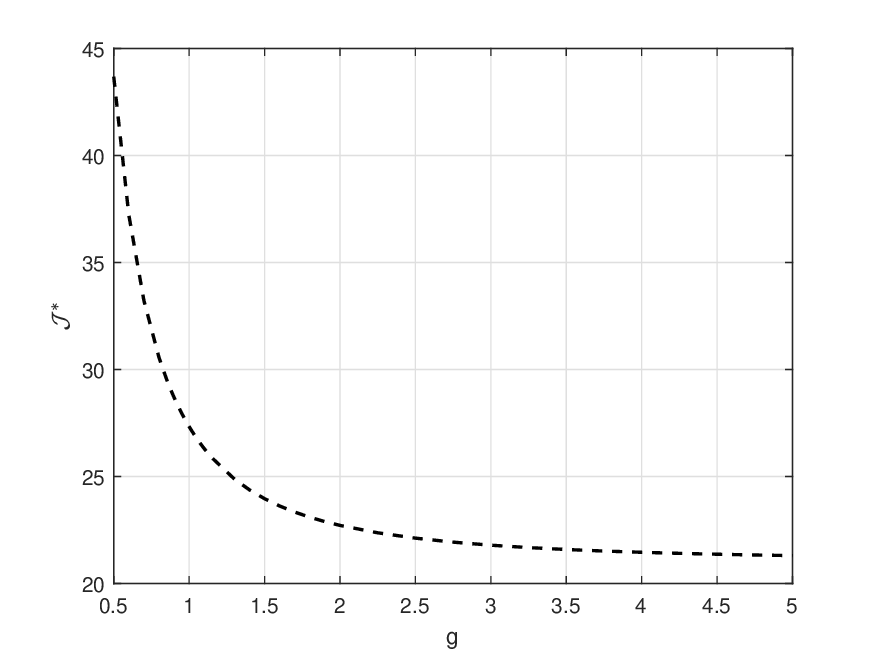}%
		\label{fig:3b}}
	\caption{(a) LQG capacity of the dynamic system \eqref{eq:30} as a function of $g$ for $J=1$. The blue curve corresponds to a fixed LQR cost constraint $p=50$, whereas the red curve corresponds to a constraint defined relative to the optimal cost, $p = \mathcal{J}^{*}+ 5$.
		(b) Optimal LQR cost as a function of \( g \), shown to aid interpretation of Fig. \ref{fig:3a}.}
	\label{fig:3}
\end{figure}

\subsection{Example}\label{sec:2-2}
Here we provide an example of LQG systems and compute its capacity. Consider the dynamic system
\begin{align}\label{eq:30}
	\mathbf{s}_{i+1}&=\begin{bmatrix} 1.4 & 0 \\ 0 & 0.4 \end{bmatrix} \mathbf{s}_i+g\begin{bmatrix} 1 \\ 1 \end{bmatrix}\mathbf{x}_i+\mathbf{w}_i \nonumber \\
	\mathbf{y}_i&= \begin{bmatrix} 1 &1 \end{bmatrix}\mathbf{s}_i+J\mathbf{x}_i+\mathbf{v}_i
\end{align}
where $g$ and $J$ are scalars, and $W=I_2$, $V=1$, and $L=0$. We also consider the LQR cost constants $Q=I_2$ and $R=1$. We illustrate the capacity for different cases of this dynamic system and different constraints $p$ below.

Figure~\ref{fig:2} shows the capacity of the dynamic system \eqref{eq:30} with $g=1$ and $J\in\{0,\,0.5,\,1\}$ as a function of $p$. Figure~\ref{fig:3a} shows the capacity for a fixed $J=1$ and varying $g$, for $p=50$ and $p=\mathcal{J}^*+5$. In Figure~\ref{fig:3b}, we plot $\mathcal{J}^*$ for varying $g$. To solve \eqref{eq:29} for this example, we used CVX \cite{cvx}.

One interesting result from Figure~\ref{fig:2} is that we can transmit information at a non-zero rate while achieving the same LQR cost as the optimal LQG controller. Another interesting observation is that the capacity is higher when $J=0$ compared to $J=1$. We believe this happens because when $J=0$, the decoder receives measurements that are not affected by the input signal (which it does not know). As a result, the decoder can estimate the states more accurately, allowing more information to be carried through the state dynamics via $G\mathbf{x}_i$ (i.e., a larger $\hat{\Sigma}$).

\section{Proof of The Main Result} \label{sec:3}	
In this section, we outline the proof of our main results by first presenting an equivalent reformulation of \eqref{eq:14}. Specifically, we rewrite the optimization problem \eqref{eq:14} as
\begin{equation}
	C_{\mathtt{LQG}}(p)\le \max_{\Upsilon\succeq 0}\frac{1}{2}\log(\frac{\det(\Psi_y)}{\det(\Psi)}) \label{eq:20}
\end{equation}
\begin{subequations}
	\begin{equation}
		\mathrm{s.t.}\;\operatorname{Tr}(\Upsilon\begin{bmatrix} K_\mathtt{LQR}^T\\I_p\end{bmatrix}\Psi_\mathtt{LQR}\begin{bmatrix} K_\mathtt{LQR}&I_p\end{bmatrix})\le p-\mathcal{J}^*, \label{eq:20c1}
	\end{equation}
	\begin{equation}
		\begin{bmatrix} F & G\\H & J\end{bmatrix}\Upsilon\begin{bmatrix} F & G\\H & J\end{bmatrix}^T-\begin{bmatrix} I_r & 0\\0 & 0\end{bmatrix}\Upsilon\begin{bmatrix} I_r & 0\\0 & 0\end{bmatrix}+\begin{bmatrix} K_p\\I_l\end{bmatrix}\Psi\begin{bmatrix} K_p^T&I_l\end{bmatrix}\succeq 0, \label{eq:20c2}
	\end{equation}
	\begin{equation}
		\Psi_y=\begin{bmatrix}H & J\end{bmatrix}\Upsilon\begin{bmatrix}H & J\end{bmatrix}^T+\Psi \label{eq:20c3}
	\end{equation}
\end{subequations}
where $(K_p,\Psi,\Sigma)$ are the Kalman filter constants from \eqref{eq:7} and \eqref{eq:8}, and $(K_\mathtt{LQR},\Psi_\mathtt{LQR},E)$ are from \eqref{eq:12}. To derive \eqref{eq:20} from \eqref{eq:14}, we define $\Upsilon=\begin{bmatrix}
	\hat{\Sigma} & \Gamma^T \\ 
	\Gamma & \Pi
\end{bmatrix}$. We use this form at certain steps in the proof to make the arguments cleaner and easier to follow.

\subsection{Overview of the Proof}

The converse is the upper bound \eqref{eq:14} of \cite{sabagICAC}. For achievability we use the time-invariant policy \eqref{eq:pol}, whose rate and cost are given by a Riccati recursion (Lemma~\ref{lemma:achTV}, known from \cite{sabagMemory,sabagICAC}). The difficulty is to show that this recursion converges to the covariance $\hat\Sigma$ of an optimizer of the convex program. This requires, at the same optimizer, the Riccati equation (Lemma~\ref{lemma:ricce}) and detectability of $(F+G\Gamma\hat\Sigma^\dagger,\,H+J\Gamma\hat\Sigma^\dagger)$ (Lemma~\ref{lemma:det}), and the latter uses $M\succ0$. We therefore impose $M\succeq\varepsilon I$ (Lemma~\ref{lemma:ach}) so that $M\succ0$ holds by construction, and we remove $\varepsilon$ at the end by Lemma~\ref{lemma:gap}.

\subsection{Proof of Theorem~\ref{thm:1}}
To prove Theorem~\ref{thm:1}, we first present the following lemma, which is the main result of our achievability proof.

\begin{lemma}[Lower Bound]\label{lemma:ach}
	\textit{Let $p>\mathcal{J}^*$. Then for every
		$\varepsilon\in(0,(p-\mathcal{J}^*)/\operatorname{Tr}(\Psi_{\mathtt{LQR}})]$ the capacity of the LQG system with LQR
		cost constraint $p$ satisfies
		\begin{equation}
			C_{\mathtt{LQG}}(p)\;\ge\; C_{\mathtt{LB},\varepsilon}(p),
		\end{equation}
		where
		\begin{align}
			C_{\mathtt{LB},\varepsilon}(p)
			&= \max_{\substack{\Gamma,\Pi,\hat{\Sigma}\\ \eqref{eq:14c1},\eqref{eq:14c2},\eqref{eq:14c3},\,\eqref{eq:14c4},\,\eqref{eq:14c5}}}
			\frac{1}{2}\log\frac{\det(\Psi_y)}{\det(\Psi)} \label{eq:ach}\\
			&\mathrm{s.t.}\quad
			\begin{bmatrix}
				\hat{\Sigma} & \Gamma^T\\
				\Gamma & \Pi-\varepsilon I
			\end{bmatrix} \succeq 0. \label{eq:ach-a}
		\end{align}
		Equivalently, \eqref{eq:ach-a} states that $M = \Pi-\Gamma\hat\Sigma^\dagger\Gamma^T \succeq \varepsilon I \succ 0$.}
\end{lemma}

The proof of this lemma is presented in Section~\ref{sec:3-2}.

\begin{remark}\label{rem:eps0}
	\textit{The restriction $\varepsilon\in\bigl(0,\,(p-\mathcal{J}^*)/\operatorname{Tr}(\Psi_{\mathtt{LQR}})\bigr]$
		in Lemma~\ref{lemma:ach} is imposed only to make the optimization problem
		\eqref{eq:ach} feasible. For larger $\varepsilon$, the constraint
		$\Pi-\Gamma\hat\Sigma^\dagger\Gamma^T\succeq\varepsilon I$ may require more input
		power than the cost budget $p-\mathcal{J}^*$ allows, and the feasible set can
		be empty. For every $\varepsilon$ in this range, the feasibility is guaranteed
		by the explicit point $\Upsilon_0=\operatorname{diag}(0,\varepsilon I_p)$, which
		satisfies all the constraints. The details are given in the proof of
		Lemma~\ref{lemma:gap}.}
\end{remark}

\begin{remark}\textit{\label{rem:ref}
		The only difference between the optimization problem \eqref{eq:14} and the one in
		Lemma~\ref{lemma:ach} is the constraint \eqref{eq:ach-a}. Constraint \eqref{eq:14c3}
		only guarantees that $\Pi-\Gamma\hat\Sigma^\dagger\Gamma^T$ is positive
		semidefinite, whereas \eqref{eq:ach-a} requires
		\begin{equation}
			M = \Pi-\Gamma\hat\Sigma^\dagger\Gamma^T \succeq \varepsilon I \succ 0.
		\end{equation}
		\\In the reformulation \eqref{eq:20}, the constraint is simply added as
		\begin{equation}
			\mathcal{S}(\Upsilon)=\Pi-\Gamma\hat\Sigma^\dagger\Gamma^T\succeq\varepsilon I.
	\end{equation}}
\end{remark}

Now, we define $\mathcal{O}_\varepsilon$ as the set of all optimal solution tuples $(\Gamma^*,\Pi^*,\hat{\Sigma}^*)$ for the optimization problem \eqref{eq:ach}. Throughout the paper, we also omit the superscript $^*$ from the optimal variables in \eqref{eq:ach} and simply denote them by $(\Gamma,\Pi,\hat{\Sigma})$. We also define $M=\Pi-\Gamma\hat{\Sigma}^{\dagger}\Gamma^T\succ 0$, which is the covariance of the i.i.d.\ part of the input policy \eqref{eq:17}.

Lemma~\ref{lemma:ach}  provides a lower bound on the capacity (Achievability). By combining this result with the upper bound in \eqref{eq:14} (Converse), we can establish Theorem~\ref{thm:1}. To this end, we present the following lemma, which shows that the gap between the upper and lower bounds can be made arbitrarily small.

\begin{lemma}[Gap between the bounds]\label{lemma:gap}
	\textit{Let $p>\mathcal{J}^*$.
		Then $\varepsilon\mapsto C_{\mathtt{LB},\varepsilon}(p)$ is nonincreasing on $(0,(p-\mathcal{J}^*)/\operatorname{Tr}(\Psi_{\mathtt{LQR}})]$, and
		\begin{equation}
			\lim_{\varepsilon\to 0^+} C_{\mathtt{LB},\varepsilon}(p) \;=\; C_{\mathtt{UB}}(p),
		\end{equation}
		where $C_{\mathtt{UB}}(p)$ denotes the upper bound from optimization problem \eqref{eq:20}.}
\end{lemma}

The proof of this lemma is presented in Appendix~\ref{app:i}.

Finally, Lemma~\ref{lemma:ach} shows that the capacity is bounded between the lower bound in \eqref{eq:ach} and the upper bound in \eqref{eq:14}. Moreover, Lemma~\ref{lemma:gap} shows that the gap between these bounds can be made arbitrarily small for any $p>\mathcal{J}^*$. Therefore, the upper and lower bounds coincide, and the capacity is given by the convex optimization problem in \eqref{eq:14} for any $p>\mathcal{J}^*$. This completes the proof of Theorem~\ref{thm:1}. \hfill $\blacksquare$

\subsection{Proof of Lemma~\ref{lemma:ach}} \label{sec:3-2}
In this section, we present the proof of Lemma~\ref{lemma:ach}. For achievability, we use the time-invariant policy \eqref{eq:17} for $i>d$, where $d$ is a finite constant to be designed, and an i.i.d. policy for $i\le d$. In short, our policy becomes:

\begin{align}
	\text{for  } i\le d&:\; \mathbf{x}_i=-K_{\mathtt{LQR}}\hat{\hat{\mathbf{s}}}_i+\mathbf{m}_i, \nonumber\\
	\text{for  } i> d&:\; \mathbf{x}_i=-K_{\mathtt{LQR}}\hat{\hat{\mathbf{s}}}_i+\Gamma\hat{\Sigma}^\dagger(\hat{\mathbf{s}}_i-\hat{\hat{\mathbf{s}}}_i)+\mathbf{m}_i. \label{eq:pol}
\end{align}

where, for $i\le d$, $\mathbf{m}_i\sim\mathcal{N}(0,\tilde{M})$, and for $i>d$, $\mathbf{m}_i\sim\mathcal{N}(0,M)$ is an i.i.d. component of the input, where $M=\Pi-\Gamma \hat{\Sigma}^\dagger \Gamma^T\succ 0$, with $(\Pi,\Gamma,\hat{\Sigma})$ being optimal variables of \eqref{eq:ach}. Also, as a reminder, $\hat{\hat{\mathbf{s}}}_i=\mathbb{E}[\hat{\mathbf{s}}_i\mid \mathbf{y}^{i-1}]$. Note that, under this policy, part of the information is transmitted through the initial inputs, while a role of the subsequent inputs is to assist the decoder in estimating them.

\begin{remark}
	\textit{$\hat{\hat{\mathbf{s}}}_i$ is a strictly causal function of $\mathbf{y}^{i-1}$ and is available to both the encoder and decoder before transmitting $\mathbf{x}_i$. Thus, the policy is causal.}
\end{remark}

This policy was proposed in \cite{sabagMemory} and \cite{sabagICAC} to achieve the conjectured lower bound. The following lemma provides the lower bound achievable using this policy, along with its corresponding LQR cost.

\begin{lemma} \label{lemma:achTV}
	\textit{Policy \eqref{eq:pol} achieves the lower bound
		\begin{equation} \label{eq:achTV}
			C_{\mathtt{LQG}}(p)\ge \lim\limits_{n\to\infty}\frac{1}{2n} \sum_{i=1}^{n} \log\frac{\det(\tilde{\Psi}_{y,i})}{\det(\Psi)},
		\end{equation}
		where $\tilde{\Psi}_{y,i},\tilde{\Sigma}_i,\tilde{K}_{y,i}$ are obtained from the following Riccati recursion:
		\begin{align} \label{eq:ll1}
			\tilde{\Sigma}_{i+1}&=(F+G\Gamma_i\hat{\Sigma}^\dagger)\tilde{\Sigma}_i (F+G\Gamma_i\hat{\Sigma}^\dagger)^T+GM_i G^T+K_p\Psi K_p^T-\tilde{K}_{y,i}\tilde{\Psi}_{y,i}\tilde{K}_{y,i}^T,\\ \label{eq:ll2}
			\tilde{\Psi}_{y,i}&=(H+J\Gamma_i\hat{\Sigma}^\dagger)\tilde{\Sigma}_i (H+J\Gamma_i\hat{\Sigma}^\dagger)^T+JM_i J^T+\Psi, \\ \label{eq:ll3}
			\tilde{K}_{y,i}&=((F+G\Gamma_i\hat{\Sigma}^\dagger)\tilde{\Sigma}_i (H+J\Gamma_i\hat{\Sigma}^\dagger)^T+GM_i J^T+K_p\Psi)\tilde{\Psi}_{y,i}^{-1}
		\end{align}
		where $\tilde{\Sigma}_1=0$, $\Gamma_i=0,M_i=\tilde{M},\Pi_i=\tilde{M}$ for $i\le d$, and $\Gamma_i=\Gamma,M_i=M,\Pi_i=\Pi$ for $i>d$. This policy also has an LQR cost of
		\begin{equation}  \label{eq:ll4}
			\mathcal{J}=\mathcal{J}^*+\lim\limits_{n\to\infty}\frac{1}{n}\sum_{i=1}^{n}[ \operatorname{Tr}(\tilde{\Sigma}_iK_\mathtt{LQR}^T\Psi_\mathtt{LQR}K_\mathtt{LQR})+\operatorname{Tr}(\Pi_i\Psi_\mathtt{LQR})+2\operatorname{Tr}(\Gamma_i\hat{\Sigma}^\dagger \tilde{\Sigma}_i K_\mathtt{LQR}^T\Psi_\mathtt{LQR})].
	\end{equation}}
\end{lemma}

The result of this lemma is presented in \cite{sabagICAC}, similar results were also previously presented in \cite{sabagfeedback} and \cite{sabagMemory}. Nevertheless, we provide a brief outline of the proof of this lemma in Appendix~\ref{app:ii}.

\begin{remark}
	\textit{Lemma~\ref{lemma:achTV} is not the contribution of this paper. The policy \eqref{eq:pol} was proposed in \cite{sabagMemory} and \cite{sabagICAC}, where the rate and cost expressions in \eqref{eq:achTV}--\eqref{eq:ll4} were derived there (see \cite[Lemmas~2--4]{sabagICAC}), following the Gaussian feedback-capacity analysis of \cite{sabagfeedback}. We include the outline proof only to keep the paper self-contained. The new ingredients of this paper are Lemmas~\ref{lemma:ach}--\ref{lemma:gap}, and Lemmas~\ref{lemma:ricce}--\ref{lemma:det}, which show that the recursion \eqref{eq:riccr} converges to a solution of \eqref{eq:ricce} that lies in the optimal set $\mathcal{O}_\varepsilon$, and the resulting proof of Theorem~\ref{thm:1}.}
\end{remark}

The Riccati recursion in Lemma~\ref{lemma:achTV} for $i> d$ becomes:

\begin{align} \label{eq:riccr}
	\tilde{\Sigma}_{i+1}&=(F+G\Gamma\hat{\Sigma}^\dagger)\tilde{\Sigma}_i (F+G\Gamma\hat{\Sigma}^\dagger)^T+GM G^T+K_p\Psi K_p^T-\tilde{K}_{y,i}\tilde{\Psi}_{y,i}\tilde{K}_{y,i}^T,\nonumber\\
	\tilde{\Psi}_{y,i}&=(H+J\Gamma\hat{\Sigma}^\dagger)\tilde{\Sigma}_i (H+J\Gamma\hat{\Sigma}^\dagger)^T+JM J^T+\Psi, \nonumber\\
	\tilde{K}_{y,i}&=((F+G\Gamma\hat{\Sigma}^\dagger)\tilde{\Sigma}_i (H+J\Gamma\hat{\Sigma}^\dagger)^T+GM J^T+K_p\Psi)\tilde{\Psi}_{y,i}^{-1}
\end{align}
where $\tilde{\Sigma}_d\succeq 0$ is a function of $\tilde{M}$ and $d$. The steady-state solution of this Riccati recursion satisfies the following Riccati equation:
\begin{align} \label{eq:ricce}
	\tilde{\Sigma}&=(F+G\Gamma\hat{\Sigma}^\dagger)\tilde{\Sigma} (F+G\Gamma\hat{\Sigma}^\dagger)^T+GM G^T+K_p\Psi K_p^T-\tilde{K}_{y}\tilde{\Psi}_{y}\tilde{K}_{y}^T,\nonumber\\
	\tilde{\Psi}_{y}&=(H+J\Gamma\hat{\Sigma}^\dagger)\tilde{\Sigma} (H+J\Gamma\hat{\Sigma}^\dagger)^T+JM J^T+\Psi, \nonumber\\
	\tilde{K}_{y}&=((F+G\Gamma\hat{\Sigma}^\dagger)\tilde{\Sigma} (H+J\Gamma\hat{\Sigma}^\dagger)^T+GM J^T+K_p\Psi)\tilde{\Psi}_{y}^{-1}
\end{align}

We now present the following lemmas to show that the lower bound in Lemma~\ref{lemma:achTV} is equivalent to the lower bound in Lemma~\ref{lemma:ach}.

\begin{lemma} \label{lemma:ricce}
	\textit{There exists a $(\Gamma,\Pi,\hat{\Sigma})\in\mathcal{O}_\varepsilon$, where $\mathcal{O}_\varepsilon$ is the set of optimal solutions to \eqref{eq:ach}, that satisfies the Riccati equation \eqref{eq:ricce}.}
\end{lemma}
Proof of this lemma is presented in Appendix~\ref{app:iii}.

The constraint \eqref{eq:14c3} in \eqref{eq:ach} is a relaxed version of the Riccati equation \eqref{eq:ricce} and is equivalent to it only if the Schur complement of the LMI \eqref{eq:14c3} is exactly zero. This relaxation makes the optimization convex. Lemma~\ref{lemma:ricce} shows that this relaxation is without loss of generality, as we can always find an optimal solution in $\mathcal{O}_\varepsilon$ that satisfies the Riccati equation.

Next, we present the following lemmas to establish the convergence of the Riccati recursion \eqref{eq:riccr} to a solution in $\mathcal{O}_\varepsilon$.

\begin{lemma} \label{lemma:help}
	\textit{Let $\mathcal{A}$ and $\mathcal{B}$ be two subspaces such that $\mathcal{A} = \mathcal{B}^\perp$. 
		Suppose there exists a matrix $K$ such that, for any vectors $a \in \mathcal{A}$ and $b \in \mathcal{B}$,
		\[
		\bar{a} K b = 0.
		\]
		Then, (i) if there exists a nonzero vector $b_1 \in \mathcal{B}$ such that $\bar{b}_1K\in  \mathcal{A}$, we can conclude that K is a singular matrix, and \\
		(ii) for any nonzero vector $a_1 \in \mathcal{A}$ and any scalar $\lambda$ that is not an eigenvalue of $K$, there exists a vector $a_2 \in \mathcal{A}$ such that
		\[
		\bar{a}_2 K a_1 \neq \lambda \bar{a}_2 a_1.
		\]}
\end{lemma}

\begin{lemma} \label{lemma:det}
	\textit{There exists a $(\Gamma,\Pi,\hat{\Sigma})\in\mathcal{O}_\varepsilon$, where $\mathcal{O}_\varepsilon$ is the set of optimal solutions to \eqref{eq:ach}, that (i) satisfies the Riccati equation \eqref{eq:ricce} and (ii) for which the pair $(F+G\Gamma\hat{\Sigma}^\dagger,H+J\Gamma\hat{\Sigma}^\dagger)$ is detectable.}
\end{lemma}

The proof of Lemma~\ref{lemma:help} is given in Appendix~\ref{app:iv}, while the proof of Lemma~\ref{lemma:det} is given in Appendix~\ref{app:v}.

Lemma~\ref{lemma:det} ensures that there exists an optimal solution that not only satisfies the Riccati equation \eqref{eq:ricce} but also satisfies the detectability condition required for convergence.

Now, from \cite{dare}, we know that the Riccati recursion \eqref{eq:riccr} converges to the steady-state solution for \eqref{eq:ricce} if (i) $\tilde{\Sigma}_d\succeq \tilde{\Sigma}$ and (ii) the pair $(F+G\Gamma\hat{\Sigma}^\dagger,H+J\Gamma\hat{\Sigma}^\dagger)$ is detectable. Under Assumption~\ref{asmp:nonzero}, the first condition can be satisfied by an appropriate choice of $d$ and $\tilde{M}$, following the argument of \cite{sabagfeedback}. Furthermore, from Lemma~\ref{lemma:det}, we know that there exists an optimal pair in $\mathcal{O}_\varepsilon$ that satisfies both the detectability condition and the Riccati equation \eqref{eq:ricce}. Thus, the Riccati recursion converges to this solution.

\begin{remark}
	\textit{The idea used here to establish the convergence of the Riccati recursion is exactly the same as that used in \cite{sabagfeedback}. However, the main contribution is to show that, in the setting with ISI, both with and without an LQR cost constraint, the detectability condition can still be satisfied while also satisfying the Riccati equation \eqref{eq:ricce}.}
\end{remark}

Now, since $\lim\limits_{i\to\infty} \tilde{\Sigma}_i=\hat{\Sigma}$, we have $\lim\limits_{i\to\infty} \tilde{\Psi}_{y,i}=\Psi_{y}$, where $\Psi_{y}$ is a variable in constraint \eqref{eq:14c4} of the optimization in Lemma~\ref{thm:1}. Therefore, the lower bound in Lemma~\ref{lemma:achTV} is exactly the lower bound in Lemma~\ref{lemma:ach}, subject to satisfying all its constraints while maintaining an LQR cost of $\mathcal{J}\le p$. This completes the proof of Lemma~\ref{lemma:ach}. \hfill $\blacksquare$

\section{Conclusion and Future Works}\label{sec:4}
In this paper, we have studied the problem of integrated communication and control. In this setup, we aim to transmit data over a dynamical system to an observer who has access to its measurements, subject to LQR cost constraints. We have computed the capacity of this channel and provided an example showing that data can be transmitted at a nonzero rate while achieving the same LQR cost (asymptotically) as the optimal control without communication. We have also discussed how feedback communication over MIMO Gaussian channels with memory and ISI is a special case of ICAC, and how the derived capacity expression applies to those channels.

There are two directions for future work. First, there is still no explicit coding scheme for general LQG systems, and finding such a scheme is an interesting direction for future research. Second, computing the capacity of the LQG system when the receiver and controller have access to different measurements is still an open problem.

\bibliographystyle{IEEEtran}
\bibliography{myref}

\appendix
\section*{Proof of Technical Lemmas}
\subsection{Proof of Lemma~\ref{lemma:gap}} \label{app:i}

Write $\Upsilon=\begin{bmatrix}\hat\Sigma & \Gamma^T\\ \Gamma & \Pi\end{bmatrix}$ and
$T=\begin{bmatrix}K_{\mathtt{LQR}}^T\\ I_p\end{bmatrix}$, and let
\[
f(\Upsilon)=\frac12\log\frac{\det\Psi_y}{\det\Psi},\qquad
\Psi_y=\begin{bmatrix}H & J\end{bmatrix}\Upsilon\begin{bmatrix}H & J\end{bmatrix}^T+\Psi .
\]
Let $\mathcal F_p$ be the feasible set of \eqref{eq:20}, i.e., all $\Upsilon\succeq 0$ satisfying
\eqref{eq:20c1} and \eqref{eq:20c2}. Then
\begin{equation*}
	C_{\mathtt{UB}}(p)=\sup_{\Upsilon\in\mathcal F_p} f(\Upsilon),
\end{equation*}
and
\begin{equation*}
	C_{\mathtt{LB},\varepsilon}(p)=\sup_{\Upsilon\in\mathcal F_p,\ \mathcal{S}(\Upsilon)\succeq \varepsilon I} f(\Upsilon),
\end{equation*}
where $\mathcal{S}(\Upsilon)=\Pi-\Gamma\hat\Sigma^\dagger\Gamma^T$.

The feasible set shrinks as $\varepsilon$ grows, so
$C_{\mathtt{LB},\varepsilon}$ is nonincreasing in $\varepsilon$, and
$C_{\mathtt{LB},\varepsilon}(p)\le C_{\mathtt{UB}}(p)$ for all $\varepsilon$. Let \[\Upsilon_0=\begin{bmatrix}
	0&0\\0&\varepsilon_0 I_p
\end{bmatrix},\]
where $\varepsilon_0=(p-\mathcal J^*)/\mathrm{Tr}(\Psi_{\mathtt{LQR}})>0$. Then $\Upsilon_0\succeq0$ and $\mathcal{S}(\Upsilon_0)=\varepsilon_0 I\succ0$. Moreover
$\operatorname{Tr}(\Upsilon_0 T\Psi_{\mathtt{LQR}}T^T)=\varepsilon_0\operatorname{Tr}(\Psi_{\mathtt{LQR}})=p-\mathcal J^*$,
so \eqref{eq:20c1} holds, and the left-hand side of \eqref{eq:20c2} equals
\[\varepsilon_0\begin{bmatrix}G\\J\end{bmatrix}\begin{bmatrix}G\\J\end{bmatrix}^T
+\begin{bmatrix}K_p\\I\end{bmatrix}\Psi\begin{bmatrix}K_p\\I\end{bmatrix}^T\succeq0,\]
because $\begin{bmatrix}I_r&0\\0&0\end{bmatrix}\Upsilon_0\begin{bmatrix}I_r&0\\0&0\end{bmatrix}=0$.
Hence $\Upsilon_0\in\mathcal F_p$.

Fix any $\Upsilon\in\mathcal F_p$ and $\delta\in(0,1]$, and set
$\Upsilon_\delta=(1-\delta)\Upsilon+\delta\Upsilon_0$.
$\mathcal F_p$ is convex, so
$\Upsilon_\delta\in\mathcal F_p$. Since $\mathcal S(\cdot)$ is matrix-concave on the PSD cone~\cite{boyd},
\begin{equation}
	\mathcal{S}(\Upsilon_\delta)\succeq(1-\delta)\mathcal{S}(\Upsilon)+\delta \mathcal{S}(\Upsilon_0)\succeq\delta\varepsilon_0 I,
\end{equation}
where we used $\mathcal{S}(\Upsilon)\succeq0$ for $\Upsilon\succeq0$. Thus $\Upsilon_\delta$ is feasible for
the $\varepsilon$-problem whenever $\varepsilon\le\delta\varepsilon_0$. Because $f$ is concave,
\begin{equation}
	f(\Upsilon_\delta)\ge(1-\delta)f(\Upsilon)+\delta f(\Upsilon_0)\ge(1-\delta)f(\Upsilon).
\end{equation}
Therefore $C_{\mathtt{LB},\varepsilon}(p)\ge(1-\delta)f(\Upsilon)$ for all $\varepsilon\le\delta\varepsilon_0$, so
\begin{equation}
	\lim_{\varepsilon\to 0^+}C_{\mathtt{LB},\varepsilon}(p)\ge(1-\delta)f(\Upsilon).
\end{equation}
Letting $\delta\to 0^+$ and then taking the supremum over $\Upsilon\in\mathcal F_p$ gives
$\lim_{\varepsilon\to 0^+}C_{\mathtt{LB},\varepsilon}(p)\ge C_{\mathtt{UB}}(p)$. Combined with the
upper bound above, this proves the claim. \hfill$\blacksquare$

\subsection{Outline of the Proof of Lemma~\ref{lemma:achTV}} \label{app:ii}

Substitute policy \eqref{eq:pol} into \eqref{eq:5}, writing $K=K_{\mathtt{LQR}}$:
\begin{align*}
	\hat{\mathbf{s}}_{i+1} &= (F+G\Gamma_i\hat\Sigma^\dagger)(\hat{\mathbf{s}}_i-\hat{\hat{\mathbf{s}}}_i)+(F-GK)\hat{\hat{\mathbf{s}}}_i+G\mathbf{m}_i+K_p\mathbf{e}_i,\\
	\mathbf{y}_i &= (H+J\Gamma_i\hat\Sigma^\dagger)(\hat{\mathbf{s}}_i-\hat{\hat{\mathbf{s}}}_i)+(H-JK)\hat{\hat{\mathbf{s}}}_i+J\mathbf{m}_i+\mathbf{e}_i .
\end{align*}

The observer's innovation is
\[
\boldsymbol{\psi}_i=\mathbf{y}_i-(H-JK)\hat{\hat{\mathbf{s}}}_i=(H+J\Gamma_i\hat\Sigma^\dagger)(\hat{\mathbf{s}}_i-\hat{\hat{\mathbf{s}}}_i)+J\mathbf{m}_i+\mathbf{e}_i,
\]
so by independence $\mathrm{cov}(\boldsymbol{\psi}_i)=\tilde\Psi_{y,i}$ as in \eqref{eq:ll2}. The innovations method gives
$\hat{\hat{\mathbf{s}}}_{i+1}=(F-GK)\hat{\hat{\mathbf{s}}}_i+\tilde K_{y,i}\boldsymbol{\psi}_i$ with gain \eqref{eq:ll3}.
Computing $\tilde\Sigma_{i+1}=\mathrm{cov}(\hat{\mathbf{s}}_{i+1}-\hat{\hat{\mathbf{s}}}_{i+1})$ gives the Riccati recursion \eqref{eq:ll1}.

Since $\boldsymbol{\psi}_i\sim\mathcal N(0,\tilde\Psi_{y,i})$ and $\mathbf{e}_i\sim\mathcal N(0,\Psi)$ are the innovations given $\mathbf{y}^{i-1}$ and $(\mathbf{x}^i,\mathbf{y}^{i-1})$, respectively,
\[
I(\mathbf{x}^n\!\to \mathbf{y}^n)=\sum_{i=1}^n\big[h(\boldsymbol{\psi}_i)-h(\mathbf{e}_i)\big]
=\tfrac12\sum_{i=1}^n\log\frac{\det\tilde\Psi_{y,i}}{\det\Psi}.
\]
Gaussian feedback achievability (Cover--Pombra, extended to \eqref{eq:1} as in \cite{sabagMemory}) shows that this directed information rate is achievable. Dividing by $n$ and letting $n\to\infty$ gives \eqref{eq:achTV}.

Split $\mathbf{s}_i=(\mathbf{s}_i-\hat{\mathbf{s}}_i)+(\hat{\mathbf{s}}_i-\hat{\hat{\mathbf{s}}}_i)+\hat{\hat{\mathbf{s}}}_i$ and use the orthogonality. Then
\begin{itemize}
	\item the term $-K\hat{\hat{\mathbf{s}}}_i$ is the optimal LQR input for the dynamics $\hat{\hat{\mathbf{s}}}_{i+1}=(F-GK)\hat{\hat{\mathbf{s}}}_i+\tilde K_{y,i}\boldsymbol{\psi}_i$, contributing $\mathrm{Tr}(\tilde K_{y,i}\tilde\Psi_{y,i}\tilde K_{y,i}^TE)$.
	\item the remaining input $\Gamma_i\hat\Sigma^\dagger(\hat{\mathbf{s}}_i-\hat{\hat{\mathbf{s}}}_i)+\mathbf{m}_i$ has covariance $\Pi_i=M_i+\Gamma_i\hat\Sigma^\dagger\Gamma_i^T$.
\end{itemize}
Replace $\tilde K_{y,i}\tilde\Psi_{y,i}\tilde K_{y,i}^T$ using \eqref{eq:ll1} and the control Riccati equation \eqref{eq:12}, exactly as in (65)--(66) of Lemma~4 in \cite{sabagICAC}. The sum telescopes, and a boundary term vanishes after dividing by $n$. This yields \eqref{eq:ll4}.

Once $\tilde\Sigma_i\to\hat\Sigma$ (Lemmas~\ref{lemma:ricce}--\ref{lemma:det}), the averages in \eqref{eq:achTV} and \eqref{eq:ll4} converge to their steady-state values. $\hfill\blacksquare$

\subsection{Proof of Lemma~\ref{lemma:ricce}} \label{app:iii}

To prove this lemma we assume that there exist no optimal solution $\Upsilon\in\mathcal{O}_\varepsilon$ satisfying Riccati equation \eqref{eq:ricce}. As a reminder $\mathcal{O}_\varepsilon$ is a set of optimal solution to the following optimization (Remark~\ref{rem:ref})
\begin{equation*}
	C_{\mathtt{LB},\varepsilon}(p)= \max_{\Upsilon\in \mathbb{R}^{(r+p)\times (r+p)}}\frac{1}{2}\log(\frac{\det(\Psi_y)}{\det(\Psi)})
\end{equation*}

\begin{align}\label{eq:c1}
	\mathrm{s.t.}\;&\operatorname{Tr}(\Upsilon\begin{bmatrix} K_\mathtt{LQR}^T\\I_p\end{bmatrix}\Psi_\mathtt{LQR}\begin{bmatrix} K_\mathtt{LQR}&I_p\end{bmatrix})\le p-\mathcal{J}^*,\;\Upsilon\succeq 0,\mathcal{S}(\Upsilon)\succeq \varepsilon I,\nonumber 
	\\ &\begin{bmatrix} F & G\\H & J\end{bmatrix}\Upsilon\begin{bmatrix} F & G\\H & J\end{bmatrix}^T-\begin{bmatrix} I_r & 0\\0 & 0\end{bmatrix}\Upsilon\begin{bmatrix} I_r & 0\\0 & 0\end{bmatrix}+\begin{bmatrix} K_p\\I_l\end{bmatrix}\Psi\begin{bmatrix} K_p^T&I_l\end{bmatrix}\succeq 0, \nonumber \\
	&\Psi_y=\begin{bmatrix}H & J\end{bmatrix}\Upsilon\begin{bmatrix}H & J\end{bmatrix}^T+\Psi.
\end{align}

We find a matrix $P$ which is is positive definite on the null space of $(H - JK_{\mathtt{LQR}})$, i.e.,
\begin{equation}
	x^T P x > 0 \quad \text{for all } x \in \ker(H - JK_{\mathtt{LQR}})
\end{equation}
and also satisfies
\begin{equation} \label{eq:c0}
	x^T (P - (F - G K_{\mathtt{LQR}})^TP(F - G K_{\mathtt{LQR}}))x > 0 \quad \text{for all } x \in \ker(H - JK_{\mathtt{LQR}}),
\end{equation}
in particular, this implies that $(P - (F - G K_{\mathtt{LQR}})^TP(F - G K_{\mathtt{LQR}}))$ is positive definite on the null space of $(H - JK_{\mathtt{LQR}})$.

Such a matrix exists because the pair $(F - G K_{\mathtt{LQR}}, H - JK_{\mathtt{LQR}})$ is detectable (Assumption~\ref{asmp:LQR}), hence $(F - G K_{\mathtt{LQR}})$ is stable on $\ker(H - JK_{\mathtt{LQR}})$. Now we define $\Xi(\Upsilon)$ as
\begin{align}\label{eq:c2}
	&\Xi(\Upsilon)=\arg\max_{\Xi\succeq 0}\operatorname{Tr}(P\Xi) \nonumber \\
	s.t\; &\begin{bmatrix} F & G\\H & J\end{bmatrix}\Upsilon\begin{bmatrix} F & G\\H & J\end{bmatrix}^T-\begin{bmatrix} I_r & 0\\0 & 0\end{bmatrix}\Upsilon\begin{bmatrix} I_r & 0\\0 & 0\end{bmatrix}+\begin{bmatrix} K_p\\I_l\end{bmatrix}\Psi\begin{bmatrix} K_p^T&I_l\end{bmatrix}\succeq  \begin{bmatrix} \Xi & 0\\0 & 0\end{bmatrix} .
\end{align}

If there were no $\Upsilon\in\mathcal{O}_\varepsilon$ satisfying \eqref{eq:ricce} we know that for all $\Upsilon\in\mathcal{O}_\varepsilon$, $\Xi(\Upsilon)$ is a non zero positive semi definite matrix. So we define $\tilde{\Upsilon}$ as

\begin{equation}
	\tilde{\Upsilon}=\arg \min_{\Upsilon\in\mathcal{O}_\varepsilon} \operatorname{Tr} (P\Xi(\Upsilon)). \label{eq:c3}
\end{equation}

We also define $\tilde{\Xi}=\Xi(\tilde{\Upsilon})$ which is non zero and positive semi definite. Consider $U=\begin{bmatrix}
	\tilde{\Xi} & -\tilde{\Xi}K_\mathtt{LQR}^T\\-K_\mathtt{LQR}\tilde{\Xi}&K_\mathtt{LQR}\tilde{\Xi}K_\mathtt{LQR}^T
\end{bmatrix}$ and $\tilde{\tilde{\Upsilon}}=\tilde{\Upsilon}+U$ which is also positive semi definite.
First we know that 
\begin{align}
	&U \begin{bmatrix} K_\mathtt{LQR}^T\\I_p\end{bmatrix}=0,\nonumber \\
	\operatorname{Tr}(&\tilde{\Upsilon}\begin{bmatrix} K_\mathtt{LQR}^T\\I_p\end{bmatrix}\Psi_\mathtt{LQR}\begin{bmatrix} K_\mathtt{LQR}&I_p\end{bmatrix})\le p-\mathcal{J}^*. \nonumber
\end{align}
So we can conclude that 
\begin{equation}\label{eq:c4}
	\operatorname{Tr}(\tilde{\tilde{\Upsilon}}\begin{bmatrix} K_\mathtt{LQR}^T\\I_p\end{bmatrix}\Psi_\mathtt{LQR}\begin{bmatrix} K_\mathtt{LQR}&I_p\end{bmatrix})\le p-\mathcal{J}^*.
\end{equation}
We can also write
\begin{align}
	&\begin{bmatrix} F & G\\H & J\end{bmatrix}\tilde{\tilde{\Upsilon}}\begin{bmatrix} F & G\\H & J\end{bmatrix}^T-\begin{bmatrix} I_r & 0\\0 & 0\end{bmatrix}\tilde{\tilde{\Upsilon}}\begin{bmatrix} I_r & 0\\0 & 0\end{bmatrix}+\begin{bmatrix} K_p\\I_l\end{bmatrix}\Psi\begin{bmatrix} K_p^T&I_l\end{bmatrix} \nonumber \\
	=&\begin{bmatrix} F & G\\H & J\end{bmatrix}{\tilde{\Upsilon}}\begin{bmatrix} F & G\\H & J\end{bmatrix}^T-\begin{bmatrix} I_r & 0\\0 & 0\end{bmatrix}{\tilde{\Upsilon}}\begin{bmatrix} I_r & 0\\0 & 0\end{bmatrix}+\begin{bmatrix} K_p\\I_l\end{bmatrix}\Psi\begin{bmatrix} K_p^T&I_l\end{bmatrix}-\begin{bmatrix} \tilde{\Xi} & 0\\0 & 0\end{bmatrix}+\begin{bmatrix} F & G\\H & J\end{bmatrix}U\begin{bmatrix} F & G\\H & J\end{bmatrix}^T. \label{eq:c5}
\end{align}
From \eqref{eq:c2} and \eqref{eq:c5} we have:
\begin{equation}
	\begin{bmatrix} F & G\\H & J\end{bmatrix}\tilde{\tilde{\Upsilon}}\begin{bmatrix} F & G\\H & J\end{bmatrix}^T-\begin{bmatrix} I_r & 0\\0 & 0\end{bmatrix}\tilde{\tilde{\Upsilon}}\begin{bmatrix} I_r & 0\\0 & 0\end{bmatrix}+\begin{bmatrix} K_p\\I_l\end{bmatrix}\Psi\begin{bmatrix} K_p^T&I_l\end{bmatrix} \succeq 0.
	\label{eq:c6}
\end{equation}
From \eqref{eq:c4} and \eqref{eq:c6} we can see that $\tilde{\tilde{\Upsilon}}$ is in feasible set of optimization \eqref{eq:c1} and From knowing $\tilde{\Upsilon}\in\mathcal{O}_\varepsilon$ and $\tilde{\tilde{\Upsilon}}\succeq\tilde{\Upsilon}$ we can conclude that 

\begin{equation}
	\label{eq:c7} \tilde{\tilde{\Upsilon}}\in\mathcal{O}_\varepsilon.
\end{equation}

Since both $\tilde{\Upsilon}$ and $\tilde{\tilde{\Upsilon}}$ belong to the optimal solution set of optimization \eqref{eq:c1}, we have
\begin{equation}
	\begin{bmatrix}H & J\end{bmatrix}\tilde{\Upsilon}\begin{bmatrix}H & J\end{bmatrix}^T+\Psi=\begin{bmatrix}H & J\end{bmatrix}\tilde{\tilde{\Upsilon}}\begin{bmatrix}H & J\end{bmatrix}^T+\Psi,
\end{equation}
which implies
\begin{align}
	\label{eq:c8}
	\begin{bmatrix}H & J\end{bmatrix}U=0,
	\\ \label{eq:c85} (H-JK_\mathtt{LQR})\tilde{\Xi}=0,
\end{align} 
so $\tilde{\Xi}$ lies in the null space of $(H-JK_\mathtt{LQR})$.

Now, using \eqref{eq:c5} and \eqref{eq:c8} we have:
\begin{align}
	&\begin{bmatrix} F & G\\H & J\end{bmatrix}\tilde{\tilde{\Upsilon}}\begin{bmatrix} F & G\\H & J\end{bmatrix}^T-\begin{bmatrix} I_r & 0\\0 & 0\end{bmatrix}\tilde{\tilde{\Upsilon}}\begin{bmatrix} I_r & 0\\0 & 0\end{bmatrix}+\begin{bmatrix} K_p\\I_l\end{bmatrix}\Psi\begin{bmatrix} K_p^T&I_l\end{bmatrix} \nonumber \\
	=&\begin{bmatrix} F & G\\H & J\end{bmatrix}{\tilde{\Upsilon}}\begin{bmatrix} F & G\\H & J\end{bmatrix}^T-\begin{bmatrix} I_r & 0\\0 & 0\end{bmatrix}{\tilde{\Upsilon}}\begin{bmatrix} I_r & 0\\0 & 0\end{bmatrix}+\begin{bmatrix} K_p\\I_l\end{bmatrix}\Psi\begin{bmatrix} K_p^T&I_l\end{bmatrix}-\begin{bmatrix} \tilde{\Xi} & 0\\0 & 0\end{bmatrix}+\begin{bmatrix} (F-GK_\mathtt{LQR})\tilde{\Xi}(F-GK_\mathtt{LQR})^T & 0\\0 & 0\end{bmatrix}. \label{eq:c9}
\end{align}
So we can say that 
\begin{equation}
	{\Xi}(\tilde{\tilde{\Upsilon}})=(F-GK_\mathtt{LQR}){\Xi}(\tilde{\Upsilon})(F-GK_\mathtt{LQR})^T. \label{eq:c10}
\end{equation}
And from \eqref{eq:c0} and \eqref{eq:c85} we can conclude that $\operatorname{Tr}(P{\Xi}(\tilde{\tilde{\Upsilon}}))<\operatorname{Tr}(P{\Xi}({\tilde{\Upsilon}}))$ which contradict \eqref{eq:c3}. So at least there exist one optimal set $\Upsilon^*\in\mathcal{O}_\varepsilon$ satisfying \eqref{eq:ricce}.

Now we only need to show that this modification does not disturb the Schur complement constraint. To this end, we only need to show that $\mathcal{S}(\tilde{\tilde{\Upsilon}})\succeq \mathcal{S}(\tilde{\Upsilon})$.

Since the Schur complement is a concave function \cite{boyd}, we can write:
\begin{align}
	\mathcal{S}(\tilde{\tilde{\Upsilon}})&= \mathcal{S}(\tilde{\Upsilon}+U) \\
	&= 2 \mathcal{S}(0.5\tilde{\Upsilon}+0.5U)\\
	&\succeq 2 (0.5\mathcal{S}(\tilde{\Upsilon})+0.5\mathcal{S}(U))\\
	&\succeq \mathcal{S}(\tilde{\Upsilon}) \succeq \varepsilon I,
\end{align}
so the Schur complement constraint also stay satisfied. \hfill
$\blacksquare$

\subsection{Proof of Lemma~\ref{lemma:help}} \label{app:iv}
To prove this lemma, we write $K$ relative to the decomposition of the space into $\mathcal{A}$ and $\mathcal{B}$ as
\begin{equation}
	K=\begin{bmatrix} K_{11} & K_{12}\\K_{21} & K_{22}\end{bmatrix},
\end{equation}
where $K_{11}:\mathcal{A}\to\mathcal{A}$, $K_{12}:\mathcal{A}\to\mathcal{B}$, $K_{21}:\mathcal{B}\to\mathcal{A}$, and $K_{22}:\mathcal{B}\to\mathcal{B}$.

For any $a \in \mathcal{A}$ and $b \in \mathcal{B}$, we have $\bar{a}Kb = 0$, hence $K_{12}=0$, and we can write $K$ as a lower triangular matrix
\begin{equation}
	K=\begin{bmatrix} K_{11} & 0\\K_{21} & K_{22}\end{bmatrix}.
\end{equation}

The eigenvalues of $K$ are therefore the eigenvalues of $K_{11}$ and $K_{22}$. Hence, if there exists a nonzero vector $b_1 \in \mathcal{B}$ such that $\bar{b}1 K \in \mathcal{A}$, then we can conclude that both $K_{22}$ and $K$ are singular matrices, which proves the first part of Lemma~\ref{lemma:help}. Also, since $\lambda$ is not an eigenvalue of $K$, it is not an eigenvalue of $K_{11}$.

Now to prove the second part of Lemma~\ref{lemma:help} we need to show that for any nonzero vector $a_1 \in \mathcal{A}$ and any scalar $\lambda$ that is not an eigenvalue of $K$, there exists a vector $a_2 \in \mathcal{A}$ such that $\bar{a}_2 K a_1 \neq \lambda \bar{a}_2 a_1$.

To show this, we set $a_2 = (K_{11} - \lambda I)a_1$, which is nonzero for any nonzero $a_1$, since $\lambda$ is not an eigenvalue of $K_{11}$. We also have $a_2 \in \mathcal{A}$. To show that $\bar{a}_2 K a_1 \neq \lambda \bar{a}_2 a_1$, we write
\begin{equation}
	\bar{a}_2(K-\lambda I)a_1=\bar{a}_2(K_{11}-\lambda I)a_1+\bar{a}_2 K_{21}a_1.
\end{equation}
Since $K_{21}a_1 \in \mathcal{B}$, we have $\bar{a}_2 K_{21}a_1=0$, and therefore
\begin{equation}
	\bar{a}_2(K-\lambda I)a_1
	=\bar{a}_2(K_{11}-\lambda I)a_1
	=\|a_2\|^2>0.
\end{equation}

Thus, we conclude that $\bar{a}_2 K a_1 \neq \lambda \bar{a}_2 a_1$.\hfill
$\blacksquare$

\subsection{Proof of Lemma~\ref{lemma:det}} \label{app:v}
From Lemma~\ref{lemma:ricce} we know that there exists an optimal set satisfying the Riccati equation (i), so we only need to show the detectability constraint. 

To show this , assume that there exist $\lambda$ and non zero $z$ such that $(F+G\Gamma\hat{\Sigma}^{\dagger})z=\lambda z$ with $|\lambda| \ge 1$, and $(H+J\Gamma\hat{\Sigma}^{\dagger})z=0$. We show that this leads to a contradiction, therefore, it cannot happen, and the pair is detectable. 

We know that any vector $z$ can uniquely written as $z=\tilde{x}+y$, where $\tilde{x}\in \operatorname{Range}(\hat{\Sigma})$ and $y\in \operatorname{ker}(\hat{\Sigma})$ (equivalently, $y \in \operatorname{ker}(\hat{\Sigma}^{\dagger})$). 

Now by setting $z=\tilde{x}+y$ we have
\begin{align} 
	&(F+G\Gamma\hat{\Sigma}^{\dagger})\tilde{x}+Fy=\lambda (\tilde{x}+y), \ \nonumber \\
	&(H+J\Gamma\hat{\Sigma}^{\dagger})\tilde{x}+Hy=0. \label{eq:a1}\end{align}

Defining $S=F+G\Gamma\hat{\Sigma}^{\dagger}-K_p(H+J\Gamma\hat{\Sigma}^{\dagger})$, it follows from \eqref{eq:a1} that
\begin{equation} \label{eq:a0}
	S(\tilde{x}+y)=\lambda (\tilde{x}+y).
\end{equation}

Note that, from the detectability of the pair $(F,H)$ (Assumption~\ref{asmp:det}), we can conclude that $\tilde{x}\neq 0$.

From Lemma~\ref{lemma:ricce} and constraints of optimization in Lemma~\ref{lemma:ach} we know that 
\begin{align}
	&\begin{bmatrix}
		\hat{\Sigma} & \Gamma^{T} \\ 
		\Gamma & \Pi
	\end{bmatrix}\succeq 0, \label{eq:36}\\
	&\hat{\Sigma}=F\hat{\Sigma}F^T+G\Pi G^T+G\Gamma F^T+F \Gamma^{T} G^T+K_p\Psi K_p^T-K_y\Psi_y K_y^{T}, \label{eq:37}\\
	&\Psi_y=H\hat{\Sigma}H^T+J\Pi J^T+H\Gamma^{T} J^T+J \Gamma H^T+\Psi, \label{eq:38}\\
	&K_y=(F\hat{\Sigma}H^T+F\Gamma^{T} J^T+G\Gamma H^T+G\Pi J^T+K_p \Psi)\Psi_y^{-1},\label{eq:39}\\ 
	&M=\Pi-\Gamma \hat{\Sigma}^\dagger \Gamma^T\succ 0.
\end{align}

Using the Schur complements of \eqref{eq:36} we have:
\begin{equation}
	\Gamma(I-\hat{\Sigma}\hat{\Sigma}^{\dagger})=0. \label{eq:39.5}
\end{equation}

Now, using \eqref{eq:38}, \eqref{eq:39}, and \eqref{eq:39.5} we can write
\begin{align}
	&(F+G\Gamma\hat{\Sigma}^{\dagger} - K_y(H+J\Gamma\hat{\Sigma}^{\dagger})) \hat{\Sigma} (F+G\Gamma\hat{\Sigma}^{\dagger} - K_y(H+J\Gamma\hat{\Sigma}^{\dagger}))^T\nonumber\\
	&=F\hat{\Sigma}F^T+F\Gamma^{T}G^T+G\Gamma F^T+G\Gamma\hat{\Sigma}^{\dagger}  \Gamma^{T}G^T\nonumber\\&+K_y(H\hat{\Sigma}H^T+J\Gamma\hat{\Sigma}^{\dagger}  \Gamma^{T}J^T+H\Gamma^{T} J^T+J \Gamma H^T)K_y^{T} \nonumber \\
	&-K_y (J\Gamma\hat{\Sigma}^{\dagger}  \Gamma^{T}G^T+J\Gamma F^T+H\hat{\Sigma}F^T+H\Gamma^{T}G^T)\nonumber\\&-(J\Gamma\hat{\Sigma}^{\dagger}  \Gamma^{T}G^T+J\Gamma F^T+H\hat{\Sigma}F^T+H\Gamma^{T}G^T)^TK_y^{T}\\
	&=F\hat{\Sigma}F^T+F\Gamma^{T}G^T+G\Gamma F^T+G\Pi G^T-GMG^T+K_y(\Psi_{y}-\Psi-JMJ^T)K_y^{T} \nonumber \\
	&-K_y(\Psi_{y}K_y^{T}-\Psi K_p^T-JMG^T)-(\Psi_{y}K_y^{T}-\Psi K_p^T-JMG^T)^TK_y^{T} \\
	&=F\hat{\Sigma}F^T+G\Pi G^T+G\Gamma F^T+F \Gamma^{T} G^T+K_p\Psi K_p^T-K_y\Psi_y K_y^{T} \nonumber \\
	&-(K_p-K_y)\Psi(K_p-K_y)^T-(K_yJ-G)M(K_yJ-G)^T. \label{eq:44}
\end{align}

Next, from \eqref{eq:37} and \eqref{eq:44} we have:
\begin{align}
	(F+G&\Gamma\hat{\Sigma}^{\dagger} - K_y(H+J\Gamma\hat{\Sigma}^{\dagger})) \hat{\Sigma} (F+G\Gamma\hat{\Sigma}^{\dagger} - K_y(H+J\Gamma\hat{\Sigma}^{\dagger}))^T\nonumber\\
	&=\hat{\Sigma}-(K_p-K_y)\Psi(K_p-K_y)^T-(K_yJ-G)M(K_yJ-G)^T. \label{eq:a2}
\end{align}

From \eqref{eq:a2}, we conclude that for any $w \in \operatorname{ker}(\hat{\Sigma})$,
\begin{align}
	\bar{w}(F+G&\Gamma\hat{\Sigma}^{\dagger} - K_y(H+J\Gamma\hat{\Sigma}^{\dagger})) \hat{\Sigma} (F+G\Gamma\hat{\Sigma}^{\dagger} - K_y(H+J\Gamma\hat{\Sigma}^{\dagger}))^Tw\nonumber\\
	&=-\bar{w}(K_p-K_y)\Psi(K_p-K_y)^Tw-\bar{w}(K_yJ-G)M(K_yJ-G)^Tw \le 0. \label{eq:a2-1}
\end{align}

Since the left-hand side of \eqref{eq:a2-1} is nonnegative and $M$ and $\Psi$ are positive definite, we conclude that
\begin{align}
	&\bar{w}K_p=\bar{w}K_y, \label{eq:a21} \\
	&\bar{w}G=\bar{w}K_yJ,\label{eq:a31} \\ 
	\bar{w}(F+G&\Gamma\hat{\Sigma}^{\dagger} - K_y(H+J\Gamma\hat{\Sigma}^{\dagger})) \hat{\Sigma}=0  \label{eq:a311}
\end{align}

From \eqref{eq:a21}, \eqref{eq:a31}, and \eqref{eq:a311} for any	$v \in \operatorname{Range}(\hat{\Sigma})$, we have
\begin{align} 
	&\bar{w}S=\bar{w}(F-K_pH), \label{eq:a41}\\
	\label{eq:a42}
	&\bar{w}Sv=0.
\end{align}

Using Lemma~\ref{lemma:help}, \eqref{eq:a41}, and \eqref{eq:a42}, together with the stability of $(F - K_p H)$ and the fact that 
$|\lambda| \ge 1$ if $y \neq 0$, we can find a $w$ such that $\bar{w}Sy \neq \lambda \bar{w}y$,
which contradicts \eqref{eq:a0}. Therefore, we conclude that $y = 0$.

Now by setting non zero $\tilde{x}=\hat{\Sigma}x$ \eqref{eq:a1} can be written as
\begin{align}
	&(F+G\Gamma\hat{\Sigma}^{\dagger})\hat{\Sigma}x=\lambda \hat{\Sigma}x, \ \nonumber \\
	&(H+J\Gamma\hat{\Sigma}^{\dagger})\hat{\Sigma}x=0. \label{eq:a11}
\end{align}

Defining $\tilde{S}=(F+G\Gamma\hat{\Sigma}^{\dagger} - K_y(H+J\Gamma\hat{\Sigma}^{\dagger})-\lambda I)$, from \eqref{eq:a2} and \eqref{eq:a11} we have:
\begin{align}
	\bar{x}\tilde{S}\hat{\Sigma} \bar{\tilde{S}}x=&	\bar{x}\hat{\Sigma}x-\bar{x}(K_p-K_y)\Psi(K_p-K_y)^Tx-\bar{x}(K_yJ-G)M(K_yJ-G)^Tx+|\lambda|^2\bar{x}\hat{\Sigma} x \nonumber\\
	-&\lambda\bar{x}\hat{\Sigma}(F+G\Gamma\hat{\Sigma}^{\dagger} - K_y(H+J\Gamma\hat{\Sigma}^{\dagger}))^Tx\nonumber\\-&\bar{\lambda}\bar{x}(F+G\Gamma\hat{\Sigma}^{\dagger} - K_y(H+J\Gamma\hat{\Sigma}^{\dagger}))\hat{\Sigma}x  \nonumber \\ =&(1-|\lambda|^2)\bar{x}\hat{\Sigma} x-\bar{x}(K_p-K_y)\Psi(K_p-K_y)^Tx-\bar{x}(K_yJ-G)M(K_yJ-G)^Tx\le 0. \label{eq:a3}
\end{align}

Since the left-hand side of \eqref{eq:a3} is non-negative, and $M$, and $\Psi$ are positive definite, we can conclude that for any $v\in\operatorname{Range}(\hat{\Sigma})$:
\begin{align}
	&|\lambda|=1, \label{eq:a4}\\
	&\bar{x}K_p=\bar{x}K_y, \label{eq:a5} \\
	&\bar{x}G=\bar{x}K_yJ,\label{eq:a6} \\
	&\bar{x}\tilde{S}v=\bar{x}(F-K_pH-\lambda I)v=0. \label{eq:a7}
\end{align}

Merging \eqref{eq:a42}, \eqref{eq:a4} and \eqref{eq:a7} using Lemma~\ref{lemma:help} we obtain that $F-K_pH$ has an eigenvalue on the unit circle, which contradicts the stability of $F-K_pH$. Therefore, \eqref{eq:a1} cannot be true, the pair $(F+G\Gamma\hat{\Sigma}^{\dagger},H+J\Gamma\hat{\Sigma}^{\dagger})$ is detectable.
\hfill$\blacksquare$

\begin{remark}
	\textit{In the case of a single-state LQG system, \eqref{eq:a5} and \eqref{eq:a6} cannot be satisfied unless $G=K_pJ$. Thus, in that case, assuming $G\neq K_pJ$ proves the Lemma~\ref{lemma:det} without the following steps, which is exactly the assumption used in \cite[Theorem 3]{sabagICAC} to compute the capacity of a single-state LQG system. However, they only proved this for SISO single-state systems, but as we can see, this condition is sufficient for any single-state MIMO LQG system.}
\end{remark}

\end{document}